\documentclass[%
aip,%
amsmath,amssymb,floatfix,
reprint,%
onecolumn,
twocolumn,%
jcp,%
longbibliography,
]{revtex4-2} 

\usepackage{graphicx}
\usepackage{dcolumn}
\usepackage{bm} 
\usepackage{color}

\usepackage[english]{babel} 

\usepackage{amsmath,amsthm,latexsym,amssymb,amsfonts,epsfig}
\usepackage{mathtools}
\usepackage{mathrsfs} 

\usepackage[outer=1.6cm, inner=1.6cm, top=2.5cm, bottom=2.5cm]{geometry}

\usepackage{ctable}

\usepackage{xcolor}
\definecolor{OliveGreen}{rgb}{0,0.6,0}

\definecolor{lightgray}{rgb}{0.83, 0.83, 0.83}

\newcommand{\vect}[1]{\boldsymbol{#1}}

\newcommand{\Intd}{\mathrm{d}}

\def\XXint#1#2#3{{\setbox0=\hbox{$#1{#2#3}{\int}$}
     \vcenter{\hbox{$#2#3$}}\kern-.5\wd0}}

\usepackage[colorlinks=true,citecolor=OliveGreen,linkcolor=OliveGreen]{hyperref}

\usepackage{epstopdf}
\usepackage{csquotes}
\usepackage{type1ec} %
\usepackage{colortbl}

\usepackage{float} 

\DeclareMathAlphabet\mathbfcal{OMS}{cmsy}{b}{n}

\allowdisplaybreaks

\emergencystretch=\maxdimen
\begin{document}


\title{Internal sites of actuation and activation in thin elastic films and membranes of finite thickness}

\author{Tyler Lutz}
\email{tlutz@uchicago.edu}
\affiliation{Institut f\"ur Physik, Otto-von-Guericke-Universit\"at Magdeburg, Universit\"atsplatz 2, 39106 Magdeburg, Germany}
\affiliation{Department of English Language and Literature, University of Chicago, Chicago, IL 60637, USA}

\author{Andreas M. Menzel}
\email{a.menzel@ovgu.de}

\affiliation{Institut f\"ur Physik, Otto-von-Guericke-Universit\"at Magdeburg, Universit\"atsplatz 2, 39106 Magdeburg, Germany}

\author{Abdallah Daddi-Moussa-Ider}
\email{abdallah.daddi-moussa-ider@open.ac.uk}

\affiliation{School of Mathematics and Statistics, The Open University, Walton Hall, Milton Keynes MK7 6AA, UK}

\date{\today}

\begin{abstract}
Functionalized thin elastic films and membranes frequently feature internal sites of net forces or stresses. These are, for instance, active sites of actuation, or rigid inclusions in a strained membrane that induce counterstress upon externally imposed deformations. 
We theoretically analyze the geometry of {isotropic}, flat, thin, linearly elastic films or membranes of finite thickness, laterally extended to infinity. At the mathematical core of such characterizations are the fundamental solutions for localized force and stress singularities associated with corresponding Green's functions. We derive such solutions in three dimensions and place them into the context of previous two-dimensional calculations.
To this end, we consider both no-slip and stress-free conditions at the top and/or bottom surfaces. We provide an understanding for why the fully free-standing thin elastic membrane leads to diverging solutions in most geometries and compare these situations to the truly two-dimensional case. A no-slip support of at least one of the surfaces stabilizes the solution, which illustrates that the divergences in the fully free-standing case are connected to global deformations. Within the mentioned framework, our results are important for associated theoretical characterizations of thin elastic films, whether supported
or free-standing, and of membranes subject to internal or external forces or stresses. 
\end{abstract}

\maketitle


\section{Introduction}

Under various circumstances, we meet thin elastic films and membranes being exposed to the action of internal forces and stresses. One obvious situation arises when such membranes are mechanically reinforced, for instance, by the inclusion of relatively firmer fibers to increase the overall stability under loading \cite{sassen2024phase}. When the overall system is stretched, mechanical counterstresses are generally exerted by these fibers on the elastic environment. 

The same is true for any rigid inclusion upon functionalization of thin elastic films or membranes. A specific example is given by elastic membranes containing rigid magnetic or magnetizable inclusions \cite{raikher2008shape}. The action of loudspeakers can be mimicked in such cases by directly exerting magnetic forces on the membrane itself, without additional external mechanical components of stimulation. 

Situations of supported elastic films on a substrate are encountered frequently during measurements of atomic force microscopy (AFM). In the field of soft matter, scales of the probes are given by lengths up to micrometers \cite{kappl2002colloidal,butt2005force}. The tip used for this purpose directly exerts forces onto the film. 
Depending on the situation, one can consider such forcing as internal,
with the AFM tip pressed into the membrane.

In a biological context, living cells generate internal and external stresses. Depending on the time scales, layers of growing and/or migrating biological cells \cite{discher2005tissue, nnetu2012impact,you2018geometry,wittmann2023collective} can be considered as thin elastic or viscoelastic films that generate internal stresses. Similarly, biofilms typically consist of living bacteria inside a thin elastic extracellular matrix \cite{flemming2010biofilm,mazza2016physics, fortune2022biofilm, fei2020nonuniform}, supported by a substrate. 

On the theoretical side, there have been some recent observations concerning two-dimensional elastic systems as representations of free-standing thin elastic membrane systems. Net forces acting in an in-plane direction within infinitely extended two-dimensional systems lead to a formal logarithmic divergence of the induced displacement field as a function of the distance from the force center \cite{phan1983image}. The two-dimensional elastic membrane, even if infinitely extended, does not offer sufficient resistance against a net force. Instead, the whole two-dimensional system will get displaced. This is in contrast to three-dimensional materials, which do not experience such diverging displacements in bulk in response to persistent point-like force centers~\cite{puljiz2019displacement}. 

The divergence in two dimensions naturally cancels if the overall force on the elastic membrane vanishes \cite{richter22}. Particularly, this is the case when so-called symmetric force dipoles \cite{hoell2019multi} act on the material. Similarly, it vanishes in the presence of, for example, a no-slip boundary \cite{menzel2017force, lutz22}. This means that, no matter how distant a boundary is, it always has a decisive influence on the displacement field resulting from a net force in the two-dimensional membrane system. 

Here, we employ a two-dimensional Fourier transformation technique to investigate thin elastic films. Yet and notably, we here do not confine ourselves to strictly two-dimensional systems. Instead, we consider elastic films and membranes of finite thickness, that is, of finite extension in the third dimension. 
The two-dimensional Fourier transformation technique has been widely utilized in elucidating the behavior of elastic sheets and deformable membranes. Its fundamental concept revolves around harnessing the translationally invariant symmetry along horizontal planes. The approach involves transformations of the partial differential equations governing membrane displacements or fluid flows — particularly in fluid dynamics — into remaining ordinary differential equations with respect to the out-of-plane coordinate.
Previously, this technique has been effectively used to describe in low-Reynolds-number hydrodynamics the behavior near walls and interfaces \cite{felderhof05, swan07, felderhof10echoing, felderhof10loss, swan10}. The same applies to the behavior of elastic membranes undergoing shear and/or bending deformation modes \cite{bickel06, felderhof06, daddi16, daddi16b, daddi16c, daddi17, daddi18jcp, daddi18epje, daddi2019axisymmetric}.

When in the present work we consider 
flat elastic membranes or thin films that are infinitely extended in the lateral direction, yet of finite thickness in the normal, third direction, the situation is intermediate between 
genuinely two-dimensional and three-dimensional, space-filling, bulk systems. An important question from the conceptual perspective is the role of force-induced divergences in this intermediate situation. How does it compare to the results for two- and three-dimensional systems?

To address this question, the remainder of the paper is structured as follows. In Sec.~\ref{sec:math}, we present the two-dimensional Fourier transformation technique applied to the governing equations of elasticity, along with solutions for various boundary conditions. Specifically, these are no-slip and/or stress-free conditions at the top and/or bottom surfaces. We explore both monopole and dipole types of singularity and address the solutions in terms of resulting Green's functions. Our results are discussed in Sec.~\ref{sec:disc}, where we also delineate the conditions under which the solutions converge and diverge. The paper concludes in Sec.~\ref{sec:concl}. 
Complex expressions pertaining to the solution are presented in tables within Appendix~\ref{appendix:tables}, while additional background concerning the calculations is included in Appendix~\ref{appendix:fourier}. 

\section{Mathematical formulation}
\label{sec:math}

The displacement field $\bm{u}(\bm{r})$ of the material points in the elastic medium is governed by the Navier-Cauchy equation
\begin{equation}
    \mu \boldsymbol{\Delta} \bm{u}(\bm{r}) + \left( \lambda + \mu \right) \boldsymbol{\nabla} \left( \boldsymbol{\nabla} \cdot \bm{u}(\bm{r}) \right) + \bm{f}(\bm{r}) = \mathbf{0} \, , 
\end{equation}
with $\mu$ and $\lambda$ denoting the shear modulus and Lam\'e parameter, respectively, and $\bm{f}(\bm{r})$ the force density field acting on the elastic material.
Both coefficients are related to each other via
\begin{equation}
    \lambda = \frac{2 \mu \nu}{1-2\nu} \, , 
\end{equation}
where $\nu \in (-1,1/2]$ is the Poisson ratio.
The components of the stress tensor read
\begin{equation}
    \sigma_{ij} = 2\mu \varepsilon_{ij} + \lambda \varepsilon_{kk} \delta_{ij} \, .
\end{equation}
In this expression, $\delta_{ij}$ represents the Kronecker delta and equal to the components of the unit matrix. $\varepsilon_{ij}=(\nabla_i u_j+\nabla_j u_i)/2$ includes the components of the linearized strain tensor.

To derive solutions for the displacement field, we utilize a two-dimensional Fourier transformation in both the $x$- and $y$-direction. This transformation is defined as follows.
The forward two-dimensional Fourier transformation of a function $g(\boldsymbol{\rho}, z)$ is denoted as $\widetilde{g}(\bm{k},z)$ and is expressed as
\begin{equation}
    \widetilde{g}(\bm{k},z) = \mathscr{F}\left\{ g(\boldsymbol{\rho},z) \right\} = \int_{\mathbb{R}^2} 
    g(\boldsymbol{\rho},z) \, e^{-i \bm{k} \cdot \boldsymbol{\rho}} \, \Intd^2 \boldsymbol{\rho} \, ,
    \label{eq:Fourier-forth}
\end{equation}
where $\bm{k}$ represents the wavevector in the two-dimensional plane of the Fourier transformation. 
Here, $\boldsymbol{\rho} = (x, y)$ denotes the projection of the position vector into the $xy$-plane.
We do not perform a Fourier transformation with respect to the~$z$-component.
Similarly, we define the inverse two-dimensional Fourier transformation as
\begin{equation}
    g(\boldsymbol{\rho},z) = \frac{1}{\left(2\pi\right)^2} \int_{\mathbb{R}^2} 
    \widetilde{g}(\bm{k}, z) \, e^{i \bm{k} \cdot \boldsymbol{\rho}} \, \Intd^2 \bm{k} \, .
    \label{eq:FTinverse}
\end{equation}
Additionally, we introduce the wavenumber $ k = |\vect{k}|$, representing the magnitude of the two-dimensional wavevector, and we define the unit vector $\hat{\vect{k}} = \vect{k}/k$.

To address the elastic equations, we adopt a cylindrical coordinate system and represent the unit wavevector as $\hat{\vect{k}} = \cos\phi \, \hat{\vect{x}} + \sin\phi \, \hat{\vect{y}}$. Additionally, we define the two-dimensional unit vector $\hat{\vect{t}}$ in the $xy$-plane perpendicular to $\hat{\vect{k}}$ such that $\hat{\vect{t}} = \sin\phi \, \hat{\vect{x}}  -\cos\phi \, \hat{\vect{y}}$.
Consequently, the Fourier-transformed displacement field and force density field can be projected onto the basis consisting of the unit vectors $\hat{\vect{k}}$ and $\hat{\vect{t}}$. 
The longitudinal and transverse components of the Fourier-transformed displacement field are defined as $\widetilde{u}_l = \widetilde{\vect{u}} \cdot \hat{\vect{k}}$ and $\widetilde{u}_t = \widetilde{\vect{u}} \cdot \hat{\vect{t}}$, respectively~\cite{bickel07}.
We refer to the $z$-component $\widetilde{u}_z$ of the Fourier-transformed displacement field as the normal component.
Similarly, the longitudinal, transverse, and normal components of the Fourier-transformed force density are denoted as $\widetilde{f}_l$, $\widetilde{f}_t$, and $\widetilde{f}_z$, respectively.

Next, in the new orthogonal basis, we derive the two-dimensional Fourier-transformed elastic equations. We find that the transverse component of the displacement field is completely independent of the longitudinal and normal components,
\begin{equation}
    \left( \partial_{zz} - k^2 \right) \widetilde{u}_t + \frac{\widetilde{f}_t}{\mu} \,  = 0 \, .
    \label{eq:transverse}
\end{equation}

The longitudinal and normal components of the displacement field are governed by a system of second-order differential equations,
\begin{subequations} \label{eq:long-und-normal}
    \begin{align}
    \left( \partial_{zz} - \frac{2 \left( 1-\nu\right)}{1-2\nu} \, k^2  \right)    \widetilde{u}_l + \frac{ik}{1-2\nu} \, \partial_z u_z +\frac{ \widetilde{f}_l }{\mu} &= 0 \, , \\[3pt]
    \left( \frac{2 \left( 1-\nu\right)}{1-2\nu} \, \partial_{zz} - k^2 \right) \widetilde{u}_z + \frac{ik}{1-2\nu} \, \partial_{z} \widetilde{u}_l 
    + \frac{ \widetilde{f}_z }{\mu} &= 0 \, .
\end{align}
\end{subequations}

Although our focus lies in seeking solutions to the elastic equations involving force singularities concentrated at one point, we remark that Eqs.~\eqref{eq:transverse} and \eqref{eq:long-und-normal} apply to any arbitrary force density $\vect{f} (\vect{r})$, given that its Fourier transformation is known.

\subsection{Solution in a bulk elastic medium}
\label{subsec:bulk}

We begin by revisiting the scenario involving a force singularity within an infinitely extended, three-dimensional, bulk elastic medium. Thus, we examine the effect of a force density represented by a Dirac delta function located at the origin of our Cartesian coordinate system, $\bm{f}(\bm{r}) = \bm{F} \, \delta^3 \left( \bm{r} \right)$. The corresponding two-dimensionally Fourier-transformed quantity, according to Eq.~(\ref{eq:Fourier-forth}), follows as $\widetilde{\bm{f}}(z) = \bm{F} \, \delta(z)$. Below, $\bm{F}$ is likewise expressed in the orthogonal basis $(\hat{\vect{k}}, \hat{\vect{t}}, \hat{\vect{z}})$.

Equation \eqref{eq:transverse} represents a second-order differential equation. The characteristic polynomial of the homogeneous equation features two distinct roots, $\pm k$. As a consequence, the solution for the transverse component of the displacement field takes the form $\widetilde{u}_t = A e^{-k|z|}$, where the constant $A$ is obtained from the jump in the first $z$-derivative due to the Dirac delta function.
Specifically,
\begin{equation}
    \left. \partial_z \widetilde{u}_t^{ \infty } \right|_{z=0^+}
    - \left. \partial_z \widetilde{u}_t^{ \infty } \right|_{z=0^-}
    = -\frac{F_t}{\mu} \, .
    \label{eq:ut}
\end{equation}
This leads to
\begin{equation}
    \widetilde{u}_t^{ \infty } = \frac{F_t}{2k} \, e^{-k|z|} \, .
    \label{eq:ut_bulk}
\end{equation}

\begin{figure}
    \centering
    \includegraphics[scale=0.45]{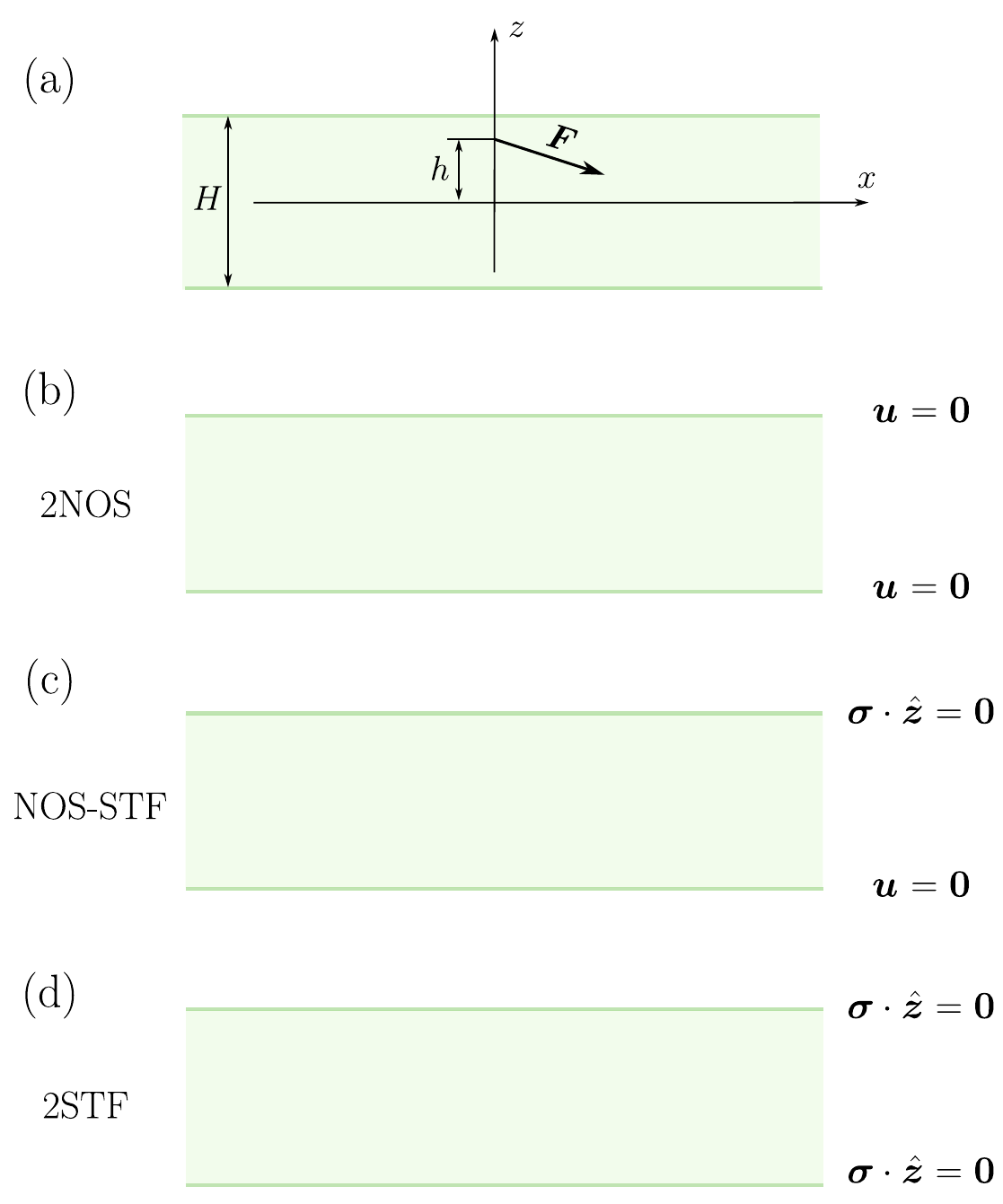}
    \caption{
    (a) Illustration of the system setup.  A force~$\bm{F}$ acts onto the thin elastic film or membrane of finite thickness~$H$ at the position $(0,0,h)$. The bottom surface of the film or membrane is located at $z=-H/2$, the top surface at $z=H/2$. For this situation, three different sets of boundary conditions are addressed. These are  
    (b)  no-slip boundary  conditions at both the top and bottom surface (2NOS),
    (c) the bottom boundary exhibiting a no-slip condition while the top boundary remains stress-free (NOS-STF),
    and (d) stress-free boundary conditions at both the top and bottom surfaces (2STF). No-slip boundary conditions imply vanishing displacements $\vect{u}=\vect{0}$, while stress-free conditions imply $\vect{\sigma}\cdot\hat{\vect{z}}=\vect{0}$.
    }
    \label{fig:memillus}
\end{figure}

The characteristic polynomials associated with the homogeneous parts of Eqs.~\eqref{eq:long-und-normal} similarly yield two distinct roots $\pm k$, each with a multiplicity of 2. Consequently, the longitudinal and normal components of the displacement field take on the forms $\widetilde{u}_l^{\infty \, \pm} = \left( \alpha_1^\pm + \alpha_2^\pm z \right) e^{\mp k z}$ and $\widetilde{u}_z^{\infty \, \pm} = \left( \beta_1^\pm + \beta_2^\pm z \right) e^{\mp k z}$, where the superscripts `plus' and `minus' correspond to the regions above and below the singularity. 
Continuity of the displacements across $z=0$ implies $\alpha_1^+ = \alpha_1^-$ and $\beta_1^+ = \beta_1^-$.
Further relations follow by introducing these functional forms into Eqs.~\eqref{eq:long-und-normal} away from the singularity position. Moreover, the Dirac delta function at $z=0$ implies the jump conditions
\begin{subequations}
    \begin{align}
    \left. \partial_z \widetilde{u}_l^{ \infty } \right|_{z=0^+}
    - \left. \partial_z \widetilde{u}_l^{ \infty } \right|_{z=0^-}
    &= -\frac{F_l}{\mu} \, , \\ 
    \left. \partial_z \widetilde{u}_z^{ \infty } \right|_{z=0^+}
    - \left. \partial_z \widetilde{u}_z^{ \infty } \right|_{z=0^-}
    &= -\frac{1-2\nu}{2 \left( 1-\nu \right)}\frac{F_z}{\mu} \, .
\end{align}
\end{subequations}
Upon solving the resulting system of linear equations, we acquire the final expressions
\begin{subequations} \label{eq:ul_uz_bulk}
    \begin{align}
        \widetilde{u}_l^{ \infty } &= \frac{e^{-k|z|}}{8\mu \left( 1-\nu\right) k}
        \left( \left( 3-4\nu - k|z| \right) F_l
        -ikz F_z \right) \, , \\
        \widetilde{u}_z^{ \infty } &= \frac{e^{-k|z|}}{8\mu \left( 1-\nu\right) k}
        \left( \left( 3-4\nu + k|z| \right) F_z
        -ikz F_l \right) \, .
    \end{align}
\end{subequations}

\subsection{Image solution technique}
\label{subsec:image}

Our goal is to quantify the static deformational response of an elastic membrane of finite thickness $H$ to an applied static force acting at position~$h \in (-H/2,H/2)$, see Fig.~\ref{fig:memillus}(a).
Initially, in the undeformed state, the membrane features two flat, parallel, infinitely extended (top and bottom) surfaces at $z=\pm H/2$. 
We examine three scenarios concerning the possible confinement of these surfaces when a force singularity is applied between these surfaces. These are (i) no-slip conditions both at the top and bottom boundaries, see Fig.~\ref{fig:memillus}(b), (ii) one no-slip condition (top boundary) and one stress-free condition (bottom boundary), see Fig.~\ref{fig:memillus}(c), and (iii) stress-free conditions both at the top and bottom boundaries, see Fig.~\ref{fig:memillus}(d).

Utilizing the bulk solution derived in Fourier space in Sec.~\ref{subsec:bulk}, we express the result as a combination of the bulk term and an additional contribution necessary to meet the prescribed boundary conditions.
This technique is commonly known as the image solution technique.
Specifically,
\begin{equation}
    \widetilde{\vect{u}} = \widetilde{\vect{u}}^\infty +\widetilde{\vect{u}}^* \, ,
\end{equation}
with $\widetilde{\vect{u}}^*$ representing image solution.

We remark that the position of the force center in the bulk elastic medium considered in Sec.~\ref{subsec:bulk} was located at the origin. Obtaining the corresponding expressions for the force acting at $z=h$ is straightforward in bulk through simple translation. To this end, $z$ is replaced by $z-h$ in the expressions for $\widetilde{u}_t$, $\widetilde{u}_l$, and $\widetilde{u}_z$ in Eqs.~\eqref{eq:ut_bulk} and \eqref{eq:ul_uz_bulk}.

The image solution $\widetilde{\vect{u}}^*$ satisfies the homogeneous parts of Eqs.~\eqref{eq:transverse} and \eqref{eq:long-und-normal}. It can be expressed in the form
\begin{equation}
    \widetilde{u}_t^* = c_1^- e^{ -k z } + c_1^+ e^{ k z } \,  \label{eq:ut_Star} 
\end{equation}
for the transverse component and
\begin{subequations} \label{eq:ul-und-uz_Star}
    \begin{align} 
    \widetilde{u}_l^* &= \left( c_2^- + c_3^- z \right) e^{ -k z }
    + \left( c_2^+ + c_3^+ z \right) e^{ k z } \, , \label{eq:ul_Star} \\
    \widetilde{u}_z^* &= \left( c_4^- + c_5^- z \right) e^{ -k z }
    + \left( c_4^+ + c_5^+ z \right) e^{ k z } \,  \label{eq:uz_Star}
\end{align}
\end{subequations}
for the longitudinal and normal components, where
\begin{subequations} \label{eq:c4c5}
    \begin{align}
    c_4^\pm &= \frac{i}{k} \left( \left( 3-4\nu\right) c_3^\pm \mp k c_2^\pm \right) \, , \\
    c_5^\pm &= \mp i c_3^\pm \, .
\end{align}
\end{subequations}
The wavenumber-dependent coefficients $c_1^\pm$, $c_2^\pm$, and $c_3^\pm$ will subsequently  be determined based on the boundary conditions specified at $z=\pm H/2$, see Fig.~\ref{fig:memillus}.

At each no-slip surface, we enforce the vanishing components of the displacement field, namely $\vect{u} = \vect{0}$, ensuring $\widetilde{u}_t = \widetilde{u}_l = \widetilde{u}_z = 0$ at these boundaries. 
For a stress-free surface, we prescribe $\boldsymbol{\sigma} \cdot \hat{\vect{z}} = \vect{0}$, implying the boundary conditions
\begin{subequations}
    \begin{align}
    \partial_z \widetilde{u}_t &= 0 \, , \\
    \partial_z \widetilde{u}_l + ik \widetilde{u}_z &= 0 \, , \\
    \left( 1-\nu\right) \partial_z \widetilde{u}_z
    +i\nu k \widetilde{u}_l &= 0 \, . \label{eq:normal-BC}
\end{align}
\end{subequations}
Particularly, we note that, if $\nu = 1/2$, Eq.~\eqref{eq:normal-BC} becomes the incompressibility condition for the elastic medium.

Subsequently, we express lengths in dimensionless units by normalizing them with respect to the thickness of the membrane $H$. Effectively we thus set a unit membrane thickness. In this way, the resulting expressions simplify.

Imposing the relevant boundary conditions on the top and bottom surfaces, we obtain a system of six linear equations for each combination of boundary conditions. Their solutions yield the expressions for the remaining unknown coefficients~$c_j^\pm$, $j\in\{1,2,3\}$.
Due to their complexity and lengthiness, and since it is straightforward to formulate the system of equations using computer algebra systems, we do not include the details of evaluation here.

First, as previously indicated, the transverse component of the displacement field, see Eq.~\eqref{eq:ut_Star}, remains unaffected by normal and longitudinal forces, see Eq.~\eqref{eq:ut}. This results in a set of equations for $c_1^\pm$ when imposing the relevant boundary conditions.
For both types of boundary conditions, the expressions for the coefficients $c_1^\pm$ follow in the form
\begin{equation}
    c_1^\pm = \frac{F_t}{2kd} \left( w_0 e^{\mp kh} + w_1 e^{k \left( 1\pm h\right)} \right) . \label{eq:c1}
\end{equation}
{Here, the specific expressions for $w_0$, $w_1$, and $d$ are determined by the specified set of boundary conditions. As depicted in Fig.~\ref{fig:memillus}, we  introduce the abbreviations 2NOS (two no-slip walls, NOS-STF (bottom no-slip and top stress-free), and 2STF (two stress-free surfaces).}
We find that $d = \mu \left( e^{2k}-1 \right) $ for 2NOS and 2STF, and $d = \mu \left( e^{2k}+1 \right)$ for NOS-STF.
In addition, $(w_0,w_1) = (1,-1)$, $(-1, \pm 1)$, and $(1,1)$ for 2NOS, NOS-STF, and 2STF, respectively.

The coefficients $c_2^\pm$ and $c_3^\pm$ entail more complex expressions, as they specify both the longitudinal and normal components of the displacement field, see Eqs.~\eqref{eq:ul-und-uz_Star} and \eqref{eq:c4c5}. This results from their combination in the coupled differential equations, see Eqs.~\eqref{eq:long-und-normal}.
For ease of reference, we introduce the abbreviations
\begin{equation}
    \sigma = 3-4\nu \, \in \, [1,7) \, , 
    \label{eq:sigma}
\end{equation}
together with $\delta_\pm = 1 \pm h$ and $\eta_\pm = 1 \pm 2h$.
Imposing the different sets of boundary conditions, the expressions for the coefficients $c_2^\pm$ and~$c_3^\pm$ determining the expressions for the longitudinal and normal components of the image displacement field, see Eqs.~\eqref{eq:ul-und-uz_Star} and \eqref{eq:c4c5}, 
for all sets of boundary conditions can be formulated as
\begin{subequations} \label{eq:c1c2}
    \begin{align}
    c_2^\pm &= \frac{1}{D}
            \sum_{n=0}^3 \left( a_n^\pm F_l + i b_n^\pm F_z \right)
            e^{k \left( n \mp (-1)^n h \right) } \, , \\
    c_3^\pm &= \frac{2k}{D}
            \sum_{n=0}^3 \left( g_n^\pm F_l + i q_n^\pm F_z \right)
            e^{ k \left( n \mp (-1)^n h \right) } \, .
\end{align}
\end{subequations}
Here, the specific expressions for the coefficients $a_n^\pm$, $b_n^\pm$, $g_n^\pm$, and $q_n^\pm$ together with the denominator~$D$ depend on the specific boundary conditions. They are provided in Tab.~\ref{tab:coefficients} of Appendix~\ref{appendix:tables}.

We now define the displacement Green's function as
\begin{equation}
    \vect{u} (\vect{r}) = \vect{ \mathcal{G } } (\vect{r}) \cdot \vect{f} \, .
\end{equation}
{It is composed of the Green's function associated with the bulk displacement field, $\vect{ \mathcal{G } }^{\infty}(\vect{r})$, and the Green's function associated with the image displacement field, $\vect{ \mathcal{G } }^{*}(\vect{r})$, }
\begin{equation}
\label{eq:Gsum}
{\vect{ \mathcal{G } } (\vect{r}) = \vect{ \mathcal{G } }^{\infty}(\vect{r}) + \vect{ \mathcal{G } }^{*}(\vect{r}).}
\end{equation}

Based on Eqs.~\eqref{eq:ut_bulk} and~\eqref{eq:ul_uz_bulk}, we deduce that in Fourier space the components of the Green's function associated with the bulk displacement field for a force acting at a distance $h$ above the center plane are
\begin{subequations} \label{eq:GreensFct_Bulk}
    \begin{align}
    \widetilde{ \mathcal{G } }_{tt}^\infty (k,z) &= \frac{1}{2k} \, e^{-k|z-h|} \, , \\
    \widetilde{ \mathcal{G } }_{ll}^\infty (k,z) &= \frac{\sigma - k|z-h|}{8\mu \left( 1-\nu\right) k} \, e^{-k|z-h|}  \, , \\ 
    \widetilde{ \mathcal{G } }_{lz}^\infty (k,z) &= -\frac{ik \left(z-h\right) }{8\mu \left( 1-\nu\right) k} \, e^{-k|z-h|} \, , \\ 
    \widetilde{ \mathcal{G } }_{zz}^\infty (k,z) &=\frac{\sigma + k|z-h|}{8\mu \left( 1-\nu\right) k} \, e^{-k|z-h|} \, .
\end{align}
\end{subequations}
These expressions can simply be read off from Eqs.~\eqref{eq:ut_bulk} and~\eqref{eq:ul_uz_bulk}, recalling the vertical shift due to the position of the force center at $z=h$ instead of $z=0$ and regarding Eq.~\eqref{eq:sigma}.
In addition, it follows that $\widetilde{ \mathcal{G } }_{zl}^\infty (k,z) = \widetilde{ \mathcal{G } }_{lz}^\infty (k,z)$.

Next, {in a similar way, we infer the the corresponding expressions for the Fourier-transformed image solution of the Green's function from the previous results for the image displacement field}.
Utilizing Eqs.~\eqref{eq:ut_Star} and~\eqref{eq:c1}, {we may express} the $tt$-component of the image Green's function 
{for all sets of boundary conditions} as
\begin{equation}
    \widetilde{ \mathcal{G } }_{tt}^* (k,z) = 
    \frac{1}{2kd} \left( w_0 \left( \psi_0^- + \psi_0^+ \right) 
    + w_1 \left( \psi_1^- + \psi_1^+ \right)
    \right) , \label{eq:Gtt_Star}
\end{equation}
where
\begin{equation}
    \psi_n^\pm = e^{ k \left( n \mp \left(-1\right)^n h \pm z \right) } \, .
\end{equation}
Moreover, it follows from Eqs.~\eqref{eq:ul-und-uz_Star}, \eqref{eq:c4c5}, and \eqref{eq:c1c2} that
\begin{equation}
    \widetilde{ \mathcal{G } }_{ij}^* (k,z) = \frac{1}{D} 
    \sum_{n=0}^3 \left( {A_{n}^-}_{ij} \psi_n^- + {A_{n}^+}_{ij} \psi_n^+ \right) \label{eq:Gij_Star}
\end{equation}
for $i,j \in \{ l,z \}$ {for all sets of boundary conditions}. 
We here defined the abbreviations
\begin{subequations}\label{eq:A}
    \begin{align}
    {A_{n}^\pm}_{ll} &= a_n^\pm + 2kz g_n^\pm \, , \\[3pt]
    {A_{n}^\pm}_{lz} &= i \left( b_n^\pm + 2kz q_n^\pm \right) , \\[3pt]
    {A_{n}^\pm}_{zl} &= \mp i \left( a_n^\pm + 2\left( kz \mp \sigma \right) g_n^\pm \right) , \\[3pt]
    {A_{n}^\pm}_{zz} &= \pm \left( b_n^\pm + 2\left( kz \mp \sigma \right) q_n^\pm \right) .
\end{align}
\end{subequations}
The expressions for $w_0$, $w_1$, $d$, $a_n^\pm$, $b_n^\pm$, $g_n^\pm$, $q_n^\pm$, and $D$ again depend on the specific set of boundary conditions and are listed below Eq.~\eqref{eq:c1} and in Tab.~\ref{tab:coefficients} of Appendix~\ref{appendix:tables}.

It is beneficial to have available the expressions of the Green's functions in the standard Cartesian coordinate system as well.
To this end, we invert the transformation outlined below Eq.~\eqref{eq:FTinverse}.
This implies 
\begin{subequations}
    \begin{align}
    \widetilde{ \mathcal{G}}_{xx}  &= \widetilde{ \mathcal{G}}_{ll}  \cos^2 \phi + \widetilde{ \mathcal{G}}_{tt}  \sin^2 \phi \, , \\
    \widetilde{ \mathcal{G}}_{yy}  &= \widetilde{ \mathcal{G}}_{ll}  \sin^2 \phi + \widetilde{ \mathcal{G}}_{tt}  \cos^2 \phi \, , \\
    \widetilde{ \mathcal{G}}_{xy} &= \left( \widetilde{ \mathcal{G}}_{ll} - \widetilde{ \mathcal{G}}_{tt} \right) \sin\phi \cos\phi \, .
\end{align}
\end{subequations}
Besides,  $ \widetilde{ \mathcal{G}}_{xz} = \widetilde{ \mathcal{G}}_{lz} \cos\phi $, $ \widetilde{ \mathcal{G}}_{yz} = \widetilde{ \mathcal{G}}_{lz} \sin\phi $, $ \widetilde{ \mathcal{G}}_{zx} = \widetilde{ \mathcal{G}}_{zl} \cos\phi $, and $ \widetilde{ \mathcal{G}}_{zy} = \widetilde{ \mathcal{G}}_{zl} \sin\phi $.
Moreover, $ \widetilde{ \mathcal{G}}_{yx} = \widetilde{ \mathcal{G}}_{xy} $, which is dictated by the geometric symmetry in the $xy$-plane.

We next proceed to the inverse Fourier transformation back to real space~\cite{bracewell99}. 
Given that the expressions in Fourier space are formulated utilizing polar coordinates $(k,\phi)$, we anticipate that the expressions in real space, after inverse Fourier transformation, can likewise be expressed in polar coordinates $(\rho, \theta)$.
Here, $\rho := \sqrt{x^2 + y^2}$ represents the radial distance from the origin, while $\theta := \arctan(y/x)$ denotes the angle formed between the positive $x$-direction and the radial direction of the considered position in space. 
A comprehensive collection of associated formulas can be found in Ref.~\onlinecite{baddour2011two}.
In Appendix~\ref{appendix:fourier}, we include an overview of the pertinent formulas utilized in the present computation.
For later reference and for the ease of computing the inverse Fourier transformation, we present the expressions of the Green's function itself in Cartesian coordinates. Conversely, its dependence on the position for convenience is maintained in polar coordinates $\rho$ and~$\theta$.

First, we list the components of the Green's function associated with the $z$-component of the applied force. A force only featuring a $z$-component is oriented perpendicular to the surfaces, implying an axisymmetric geometry. The inverse Fourier transformation yields
\begin{align}
  \mathcal{ G}_{\rho z}  &= \frac{i}{2\pi} \int_{0}^{\infty} \widetilde{ \mathcal{G} }_{lz} (k,z) J_1 (\rho k) \, k \, \Intd k \, , 
  \label{eq:Grhoz} \\[3pt]
  \mathcal{ G}_{zz} &= \frac{1}{2\pi}
  \int_{0}^{\infty}  \widetilde{ \mathcal{G} }_{zz} (k,z) J_0 (\rho k) \, k \, \Intd k \, . \label{eq:Gzz}
\end{align}
In our notation, $J_n$ refers to the Bessel function of the first kind of order $n$~\cite{abramowitz72}.
It follows that $\mathcal{G }_{xz}=\mathcal{G }_{\rho z}\cos\theta$ and $\mathcal{G }_{yz}=\mathcal{ G}_{\rho z}\sin\theta$ in Cartesian coordinates.

The {remaining components of the Green's function are calculated in analogy to that.} 
We introduce the notation
\begin{equation}
 \widetilde{ \mathcal{G} }^{\pm}_n (k, \rho, z) := \frac{k}{4\pi} \left( \widetilde{ \mathcal{G} }_{tt}(k,z) \pm \widetilde{ \mathcal{G} }_{ll}(k,z) \right) J_n (\rho k) \, .
\end{equation}
Then, we obtain
\begin{subequations}
\begin{align}
\mathcal{G}_{xx }  &=  \int_0^{\infty} \left( \widetilde{ \mathcal{G}}^{+}_0 (k,\rho, z) +   \widetilde{ \mathcal{G} }^{-}_2 (k,\rho, z) \cos \left( 2\theta \right) \right)  \Intd k \, ,  \\[3pt]
\mathcal{G }_{yy}  &=  \int_0^{\infty} \left( \widetilde{ \mathcal{G} }^{+}_0 (k,\rho,z) - \widetilde{ \mathcal{G} }^{-}_2 (k,\rho,z) \cos \left( 2\theta \right) \right) \Intd k \, .
\end{align}
\end{subequations}
Moreover,
\begin{equation}
    \mathcal{ G}_{xy} = \int_0^\infty \mathcal{ G}^{-}_2 (k,\rho, z)  \sin \left( 2\theta \right) \, \Intd k \, .
\end{equation}
Again, $\mathcal{ G}_{yx}=\mathcal{ G}_{xy}$, which agrees with the geometric symmetry within the $xy$-plane.
Furthermore, $\mathcal{G }_{zx}=\mathcal{ G}_{z\rho}\cos\theta$ and $\mathcal{G }_{zy}=\mathcal{ G}_{z\rho}\sin\theta$, where
\begin{equation}    
    \mathcal{ G}_{z\rho}  = \frac{i}{2\pi} \int_{0}^{\infty} \widetilde{ \mathcal{G} }_{zl} (k,z) J_1 (\rho k) \, k \, \Intd k \, .
    \label{eq:Gzrho}
\end{equation}

In a bulk elastic medium, which is infinitely extended in all space directions, the Green's function reads\cite{stakgold2011green}
\begin{equation}
    \widetilde{ \mathcal{G} }_{ij}^\infty = \frac{1}{16\pi \mu \left(1-\nu \right)}
    \left( \frac{\sigma}{r} \, \delta_{ij} + \frac{ r_i r_j }{r^3} \right) , 
    \label{eq:green-infinity}
\end{equation}
where $r := |\vect{r}| = \sqrt{\rho^2 + \left( z-h\right)^2}$ denotes the distance from the position of the applied force singularity.
The result in Eq.~(\ref{eq:green-infinity}) is obtained by inverse Fourier transformation of Eqs.~\eqref{eq:GreensFct_Bulk}.
For $\nu=1/2$, implying $\sigma =1$ {in Eq.~\eqref{eq:sigma}}, representing an incompressible elastic medium that does not allow any divergence of the displacement field at any position in space, we recover the classical solution recognized in fluid mechanics as the Oseen tensor \cite{happel12, kim13}. This tensor characterizes the behavior of an incompressible fluid under the influence of a force singularity under low-Reynolds-number conditions. There, $\mu$ assumes the role of the (dynamic) shear viscosity {and $\vect{u}(\vect{r})$ takes the role of the flow field of the fluid}.

\subsection{Convergence of the Green's functions {for force monopoles}}
\label{sec:convergence}

{We here consider the displacement fields arising in a laterally infinitely extended elastic film or membrane of finite thickness. The expressions for the Green's functions quantifying these displacements  in real space in response to the localized application of a force were derived in Sec.~\ref{subsec:image}, see Eqs.~\eqref{eq:Grhoz}--\eqref{eq:Gzrho}.} They are provided by integral expressions that result from the inverse two-dimensional Fourier transformation back to real space, {see Eq.\eqref{eq:FTinverse}}. We discuss the divergence of these expressions for the different types of boundary conditions. 

Considering that the Green's functions exhibit a screened exponential decay as the wavenumber $k$ approaches infinity, which results from the Bessel functions {in Eqs.~\eqref{eq:Grhoz}--\eqref{eq:Gzrho}}, problems of convergence {of the integrals} do not arise when $k$ approaches infinity. Our primary concern remains the convergence of the {integrals determining the} Green's functions {in Eqs.~\eqref{eq:Grhoz}--\eqref{eq:Gzrho}} as $k$ tends towards zero.
Near $ k=0 $, the Bessel functions $ J_n(\rho k) $ can be approximated as $ \left( \rho k/2 \right)^n/n! + \mathcal{O} \left( k^{n+2} \right) $. Thus, $ \widetilde{\mathcal{G}} $ must exhibit $ \mathcal{O} \left( k^{-1}\right)$ or higher-order behavior in $k$ {near $k=0$ to warrant convergence of the integrals}.
{The cause lies with the factor $k$ that the Green's functions $ \widetilde{\mathcal{G}} $ are multiplied with in polar coordinates when performing the two-dimensional inverse Fourier transformation in Eq.~\eqref{eq:FTinverse}. Overall, this requires} an exponent of $k$ of at least $-1$ {in $ \widetilde{\mathcal{G}} $} to hinder divergence of the integrals at $ k=0$.

For boundary conditions {involving at least one no-slip surface, that is, for} the categories 2NOS and NOS-STF, we observe that the diagonal components $\widetilde{\mathcal{G}}_{tt}$, $\widetilde{\mathcal{G}}_{ll}$, and $\widetilde{\mathcal{G}}_{zz}$ all exhibit behavior of the order $\mathcal{O} \left( k^{-1}\right)$, {see Eqs.~\eqref{eq:green-infinity}--\eqref{eq:A} together with the expressions of the parameters listed in Tab.~\ref{tab:coefficients} of Appendix~\ref{appendix:tables}. Moreover,}  the off-diagonal components $\widetilde{\mathcal{G}}_{lz}$ and $\widetilde{\mathcal{G}}_{zl}$ show behavior on the order of $\mathcal{O} \left( 1 \right)$, {as equally inferred from this set of equations. Overall,} this indicates convergence of the Green's function {for all cases involving at least one no-slip boundary}.

{The situation becomes markedly different in the case of two stress-free boundaries (2STF). In this situation, from Eqs.~\eqref{eq:Gtt_Star}--\eqref{eq:A} together with the coefficients listed in Tab.~\ref{tab:coefficients} in Appendix~\ref{appendix:tables}}, we observe that $\widetilde{\mathcal{G}}_{tt}$ and $\widetilde{\mathcal{G}}_{ll}$ exhibit behavior on the order of $\mathcal{O} \left( k^{-2} \right)$. Even more so, $\widetilde{\mathcal{G}}_{lz}$ and $\widetilde{\mathcal{G}}_{zl}$ show behavior of the order of $\mathcal{O} \left( k^{-3} \right)$, and $\widetilde{\mathcal{G}}_{zz}$ is of the order of $\mathcal{O} \left( k^{-4} \right)$. These relations indicate a non-convergent Green's function {in real space when the integrals in Eqs.~\eqref{eq:Grhoz}--\eqref{eq:Gzrho} are carried out}.

\subsection{{Convergence of the Green's functions for force dipoles}}
\label{sec:convergence-dipoles}

Having analyzed the behavior in response to  singularities consisting of force monopoles, our focus shifts to singularities of force dipoles. Again, we investigate how the behavior varies for the different boundary conditions enforced at the surfaces.

First, we introduce the $\hat{\vect{a}}$-directed Green's function as 
\begin{equation}
    \vect{G} (\vect{r} 
    ; \hat{\vect{a}}) = \vect{\mathcal{G}} (\vect{r}, {\vect{r}_0}) \cdot \hat{\vect{a}} \, ,
    \label{eq:Gadirected}
\end{equation}
where $\bm{a}$ is a unit vector {and $\vect{r}_0$ is the position of applying the force singularity.}
Next, we calculate the gradient {$\boldsymbol{\nabla}_0$} of the $\hat{\vect{a}}$-directed Green's function with respect to the {position $\vect{r}_0$}. In this way, we define the Green's function for the force dipole {composed of $\hat{\vect{a}}$-directed forces}  as
\begin{equation}
    \vect{G}_\mathrm{D} (\vect{r} 
    ; \hat{\vect{a}}, \hat{\vect{b}} )
    = 8\pi \mu \left( \hat{\vect{b}} \cdot \boldsymbol{\nabla}_0 \right)
    \vect{G} (\vect{r} 
    ; \hat{\vect{a}}) \, .
    \label{eq:GD_diff}
\end{equation}
{This force dipole extends along the direction given by the unit vector $\bm{b}$.}

{
In this expression, the differentiation $\boldsymbol{\nabla}_0(\cdot)$ with respect to the position $\vect{r}_0$ of the force singularity  stems from casting the discrete setup of two nearby force centers into a differential form. During this process, as a limit, the distance between the two centers of the point forces tends to zero. This is how we turn from the two infinitesimally separated force centers to the force dipole.  
In contrast to $\boldsymbol{\nabla}_0(\cdot)$, the regular $\boldsymbol{\nabla}(\cdot)$ is the differentiation with respect to the regular space coordinate. Due to the translational symmetry along the $xy$ plane, it follows from Eq.~\eqref{eq:GD_diff} in Cartesian coordinates that $\partial / \partial x_0 = -\partial / \partial x$ and $\partial / \partial y_0 = -\partial / \partial y$. 
However, in the presence of a boundary, $\partial / \partial h$ does not equate to $-\partial /\partial z$. In fact, $\vect{\mathcal{G}} (\vect{r}, {\vect{r}_0})$ in Eq.~\eqref{eq:Gadirected} depends on $x-x_0$ and $y-y_0$, but not on $z-z_0$, where $\vect{r}_0=(x_0,y_0,z_0)$. 
}

Next, we define the stresslet as the symmetric part of the force dipole, expressed as
\begin{equation}
    \vect{G}_\mathrm{S} (\vect{r}; \hat{\vect{a}}, \hat{\vect{b}} ) = 
    \frac{1}{2} 
    \left( \vect{G}_\mathrm{D} (\vect{r}; \hat{\vect{b}}, \hat{\vect{a}} ) 
    + \vect{G}_\mathrm{D} (\vect{r}; \hat{\vect{a}}, \hat{\vect{b}} )
    \right) \, .
    \label{eq:GS}
\end{equation}
Likewise, we define the rotlet as the antisymmetric part,
\begin{equation}
    \vect{G}_\mathrm{R} (\vect{r}; \hat{\vect{c}} ) = 
    \frac{1}{2} \left( \vect{G}_\mathrm{D} (\vect{r}; \hat{\vect{b}}, \hat{\vect{a}} ) - \vect{G}_\mathrm{D} (\vect{r}; \hat{\vect{a}}, \hat{\vect{b}} ) \right) \, ,
    \label{eq:GR}
\end{equation}
where $\hat{\vect{c}} = \hat{\vect{a}} \times \hat{\vect{b}}$.

It is more convenient to derive the expressions of the force dipole in Fourier space {according to Eq.~\eqref{eq:Fourier-forth}}. Then, we obtain
\begin{subequations}
    \label{eq:GD-FT}
    \begin{align}
    \widetilde{\vect{G}}_\mathrm{D} (\vect{k}, z; \hat{\vect{a}}, \hat{\vect{x}} ) &=
    -ik \cos\phi \,\, \widetilde{\vect{G}} (\vect{k}, z; \hat{\vect{a}}) \, , 
    \label{eq:GDx} \\
    \widetilde{\vect{G}}_\mathrm{D} (\vect{k}, z; \hat{\vect{a}}, \hat{\vect{y}} ) &=
    -ik \sin\phi \,\, \widetilde{\vect{G}} (\vect{k}, z; \hat{\vect{a}}) \, , 
    \label{eq:GDy} \\
     \widetilde{\vect{G}}_\mathrm{D} (\vect{k}, z; \hat{\vect{a}}, \hat{\vect{z}} ) &=
     \partial_h \widetilde{\vect{G}} (\vect{k}, z; \hat{\vect{a}}) \, ,
    \end{align}
\end{subequations}
{recalling that the Fourier transformation is performed only in two dimensions so that the arguments are given by the two-dimensional wavevector $\vect{k}$ together with the real-space $z$ coordinate. The minus signs in Eqs.~\eqref{eq:GDx} and \eqref{eq:GDy} result from the relations $\partial / \partial x_0 = -\partial / \partial x$ and $\partial / \partial y_0 = -\partial / \partial y$ noted above.}

When turning to the inverse {two-dimensional} Fourier transformation {according to Eq.~\eqref{eq:FTinverse}}, the solution is expressed in integral form
\begin{equation}
    \vect{G}_\mathrm{D} (\vect{r}; \hat{\vect{a}}, \hat{\vect{b}})
    = \int_0^\infty k \, \boldsymbol{\Gamma} ( \vect{r}, k ; \hat{\vect{a}}, \hat{\vect{b}}) \, \Intd k \, . \label{eq:GD_Gamma}
\end{equation}
Integration with respect to the variable $\phi$ has already been performed {in the function $\boldsymbol{\Gamma} ( \vect{r}, k ; \hat{\vect{a}}, \hat{\vect{b}})$. This calculation can be supported by conventional computer algebra systems. In Cartesian coordinates, the expression for $\boldsymbol{\Gamma} ( \vect{r}, k ; \hat{\vect{a}}, \hat{\vect{b}})$ needs to be evaluated with respect to all $\hat{\vect{a}}, \hat{\vect{b}}\in\{\hat{\vect{x}},\hat{\vect{y}},\hat{\vect{z}}\}$. Each such combination of $\hat{\vect{a}}$ and $ \hat{\vect{b}}$ implies three Cartesian components of $\vect{G}_\mathrm{D} (\vect{r}; \hat{\vect{a}}, \hat{\vect{b}})$ in Eq.~\eqref{eq:GD_Gamma}.}
We present the corresponding components of $\vect{G}_\mathrm{D} (\vect{r}; \hat{\vect{a}}, \hat{\vect{b}})$ 
in Tab.~\ref{tab:Gamma} of Appendix~\ref{appendix:tables}.

Next, we investigate the convergence of the integral {in Eq.~\eqref{eq:GD_Gamma} in analogy to the procedure outlined in Sec.~\ref{sec:convergence}. To this end,} we examine whether $\boldsymbol{\Gamma}$ is of order $  k^{-1} $ or higher as $k$ approaches~$0$. {From the relations in Eqs.~\eqref{eq:Gadirected} and \eqref{eq:GD-FT}, we infer convergence for $\vect{G}_\mathrm{D} $ whenever the integrals for $\vect{\mathcal{G}}$ converge. We found convergence for the pure force singularity in all cases that involve at least one no-slip boundary (2NOS and NOS-STF). Thus,} we may assume converge of the solution under these boundary conditions also for $\vect{G}_\mathrm{D} $. 

Therefore, the remaining task is to consider the situation of two stress-free boundaries (2STF).
It turns out that in this case the solution for the force dipole generally can still show divergence. For certain geometries, it converges while the solution for the pure force singularity diverges.

\begin{table}
    {\renewcommand{\arraystretch}{2.}
\begin{tabular}{|c|c|c|}
    \cline{2-3}
    \multicolumn{1}{c|}{} & $\boldsymbol{\Gamma} ( \vect{r}, k ; \hat{\vect{a}}, \hat{\vect{b}})$ & 
    ~~Expression for 2STF up to $ \mathcal{O} \left( k^{-3} \right) $~~ \\
    \hline
    ~~1~~ & $\boldsymbol{\Gamma} ( \vect{r}, k ; \hat{\vect{x}}, \hat{\vect{x}})$  & $12 \left(1-\nu\right) h k^{-2} \, \hat{\bm{z}} $ \\
   2 & ~~~$\boldsymbol{\Gamma} ( \vect{r}, k ; \hat{\vect{z}}, \hat{\vect{x}})$~~~ & ~~~$ 12  \left(1-\nu\right) k^{-2} \left( -z \, \hat{\bm{x}} + x \, \hat{\bm{z}} \right) $~~~ \\
    \hline
    3 & $\boldsymbol{\Gamma} ( \vect{r}, k ; \hat{\vect{y}}, \hat{\vect{y}})$ & $ 12 \left(1-\nu\right) h k^{-2} \, \hat{\bm{z}} $ \\
    4 & $\boldsymbol{\Gamma} ( \vect{r}, k ; \hat{\vect{z}}, \hat{\vect{y}})$ & $ 12 \left(1-\nu\right) k^{-2} \left( -z \, \hat{\bm{y}} + y \, \hat{\bm{z}} \right) $ \\
    \hline
    5 & $\boldsymbol{\Gamma} ( \vect{r}, k ; \hat{\vect{x}}, \hat{\vect{z}})$ & $ 12 \left(1-\nu\right) k^{-2} \left( z \, \hat{\bm{x}} - x \, \hat{\bm{z}} \right) $ \\
    6 & $\boldsymbol{\Gamma} ( \vect{r}, k ; \hat{\vect{y}}, \hat{\vect{z}})$ & $ 12 \left(1-\nu\right) k^{-2} \left( z \, \hat{\bm{y}} - y \, \hat{\bm{z}} \right) $ \\
    7 & $\boldsymbol{\Gamma} ( \vect{r}, k ; \hat{\vect{z}}, \hat{\vect{z}})$ & $-24 \, \nu h k^{-2} \,  \hat{\bm{z}}$ \\
    \hline
\end{tabular}
}
\caption{{Leading-order expressions in $k$} of the function {$\boldsymbol{\Gamma}(\vect{r},k;\hat{\vect{a}},\hat{\vect{b}})$ around $k=0$ with $\hat{\vect{a}},\hat{\vect{b}}\in\{\hat{\vect{x}},\hat{\vect{y}},\hat{\vect{z}}\}$} up to $\mathcal{O} \left( k^{-3} \right)$. {The function $\vect{\Gamma}$ determines the Green's function $\vect{G}_D$ for the force dipole, see Eq.~\eqref{eq:GD_Gamma}}. Here, we refer to the scenario of a free-standing elastic film or membrane of finite thickness featuring two parallel surfaces of stress-free boundary conditions (2STF). 
The combinations $\boldsymbol{\Gamma} ( \vect{r}, k ; \hat{\vect{x}}, \hat{\vect{y}})$ and $\boldsymbol{\Gamma} ( \vect{r}, k ; \hat{\vect{y}}, \hat{\vect{x}})$ are omitted because both are of $\mathcal{O} \left( 1 \right)$ to leading order in $k$ at $k=0$ and thus imply convergent integrals in Eq.~\eqref{eq:GD_Gamma}.
}
\label{tab:Gamma_smallK}
\end{table}

The leading-order behavior of $\vect{\Gamma}$ with respect to $k$ near $k=0$ can be determined from the expressions in Tab.~\ref{tab:Gamma} in Appendix~\ref{appendix:tables} systematically using computer algebra.
Upon examination, we observe that only the configurations $\vect{\Gamma}(\vect{r},k;\hat{\vect{x}},\hat{\vect{y}})$  and $\vect{\Gamma}(\vect{r},k;\hat{\vect{y}},\hat{\vect{x}})$ satisfy the criteria necessary for the integral in Eq.~\eqref{eq:GD_Gamma} to generally converge. Thus, only the Green's functions $\vect{G}_\mathrm{D} ( \vect{r} ; \hat{\vect{x}}, \hat{\vect{y}})$ and $\vect{G}_\mathrm{D} ( \vect{r} ; \hat{\vect{y}}, \hat{\vect{x}})$ for the force dipole are well defined in general.
We present in Tab.~\ref{tab:Gamma_smallK} the leading-order terms in the series expansion {of the remaining combinations of $\boldsymbol{\Gamma}(\vect{r},k;\hat{\vect{a}},\hat{\vect{b}})$ around $k=0$ with $\hat{\vect{a}},\hat{\vect{b}}\in\{\hat{\vect{x}},\hat{\vect{y}},\hat{\vect{z}}\}$. From there, it is obvious that no other combination of $\hat{\vect{a}}$ and $\hat{\vect{b}}$ provides the required leading-order behavior of $k^{-1}$ or higher so that convergence of the integral in Eq.~\eqref{eq:GD_Gamma} does not generally result.}

Concerning the Green's function for the stresslet $\vect{G}_\mathrm{D}$, see Eq.~\eqref{eq:GS}, it is evident that  convergence is achieved, nevertheless, if the stresslet is positioned precisely at the center plane of the elastic film or membrane. This implies $h=0$. Then, we find from Tab.~\ref{tab:Gamma_smallK} that the leading orders in $k^{-2}$ cancel for $\boldsymbol{\Gamma}(\vect{r},k;\hat{\vect{x}},\hat{\vect{x}})$, $\boldsymbol{\Gamma}(\vect{r},k;\hat{\vect{y}},\hat{\vect{y}})$, and $\boldsymbol{\Gamma}(\vect{r},k;\hat{\vect{z}},\hat{\vect{z}})$. Moreover, upon symmetrization in Eq.~\eqref{eq:GS}, the leading orders in $k^{-2}$ cancel for $\boldsymbol{\Gamma}(\vect{r},k;\hat{\vect{x}},\hat{\vect{z}})+\boldsymbol{\Gamma}(\vect{r},k;\hat{\vect{z}},\hat{\vect{x}})$ and $\boldsymbol{\Gamma}(\vect{r},k;\hat{\vect{y}},\hat{\vect{z}})+\boldsymbol{\Gamma}(\vect{r},k;\hat{\vect{z}},\hat{\vect{y}})$. Therefore, the leading order is shifted to at least $k^{-1}$ around $k=0$, thus satisfying the required criterion of convergence for the integral in Eq.~\eqref{eq:GD_Gamma}.
Besides, this implies that for the Green's function of the rotlet {$\vect{G}_\mathrm{R}$, see Eq.~\eqref{eq:GR},} only the configuration of $\hat{\vect{c}}= \hat{\vect{z}}$ implies general convergence from the integrals in Eq.~\eqref{eq:GD_Gamma}.

\section{Discussion}
\label{sec:disc}

We touched here the subtle question of possible divergences in the displacements of elastic films and membranes when exposed to mechanical force density fields. Conversely to many previous works, we here focused on systems of finite, non-vanishing thickness. In deriving associated Green's functions, we concentrated on the response to imposed forces, {see Secs.~\ref{subsec:bulk}--\ref{sec:convergence}}, stresslets (symmetric contributions to force dipoles), see Sec.~\ref{sec:convergence-dipoles} and rotlets implying torques (antisymmetric contributions to force dipoles), see Sec.~\ref{sec:convergence-dipoles} as well. 

During these investigations, we encountered possible divergences in the resulting displacement fields. Generally, we found that the presence of at least one no-slip boundary stabilizes the system so that no divergence is observed, {see Sec.~\ref{sec:convergence}}. However, divergences in the displacement field emerge for free-standing elastic membranes that are infinitely extended to the lateral sides and of finite thickness in the normal direction. That is, stress-free boundary conditions apply on both the top and bottom surface. The divergences are connected to small wavenumbers and thus large length scales, {see Sec.~\ref{sec:convergence}}. In fully three-dimensional, bulk systems such divergences are not observed \cite{puljiz2019displacement}. 

We first compare to the situation of two-dimensional elastic systems, that is, infinitely thin, flat elastic membranes. In that case, logarithmic divergences are found for displacement fields in response to net forces \cite{phan-thien87}, but not in response to force dipoles \cite{richter22}, all applied along in-plane directions. 

In our considerations, the two-dimensional situation is recovered when applying forces and force dipoles that are oriented in the $xy$-plane and positioned in the center plane at $z=0$ ($h=0$), {see Fig.~\ref{fig:memillus}}. In that case, in strictly two dimensions, net forces lead to diverging displacement fields, but force dipoles do not \cite{richter22}. Indeed, we recover this convergence for force dipoles, if $\vect{G}_\mathrm{D} (\vect{r}; \hat{\vect{x}}, \hat{\vect{x}})$ and $\vect{G}_\mathrm{D} (\vect{r}; \hat{\vect{y}}, \hat{\vect{y}})$ are evaluated for positioning the force dipoles in the center plane. 
This is observed from Eq.~\eqref{eq:GD_Gamma} together with entries 1 and 3 in Tab.~\ref{tab:Gamma_smallK}. For $h=0$, the leading order in $k$ around $k=0$ for  $\vect{\Gamma} (\vect{r},k; \hat{\vect{x}}, \hat{\vect{x}})$ and $\vect{\Gamma} (\vect{r},k; \hat{\vect{y}}, \hat{\vect{y}})$ shifts to at least $k^{-1}$, thus ensuring convergence of the integral in Eq.~\eqref{eq:GD_Gamma}, see also Sec.~\ref{sec:convergence-dipoles}. Along the same lines, from the orders of $\vect{\Gamma}$ mentioned in the caption of Tab.~\ref{tab:Gamma_smallK}, we infer that the expressions of  $\vect{G}_\mathrm{D} (\vect{r}; \hat{\vect{y}}, \hat{\vect{x}})$ and $\vect{G}_\mathrm{D} (\vect{r}; \hat{\vect{y}}, \hat{\vect{x}})$ are always well defined. Therefore, the symmetric positioning of force dipoles at $h=0$ recovers the situation of strictly two-dimensional considerations \cite{richter22}, where all resulting displacement fields remain finite.
Conversely, off-center positioning of force dipoles, {implying $h\neq0$,} can lead to diverging displacement fields, signaled by the nonconverging integral in Eq.~\eqref{eq:GD_Gamma} when $k$ approaches $0$ as highlighted by the expressions in Tab.~\ref{tab:Gamma_smallK}.

Additionally, we note that displacements according to stresslets corresponding to the Green's functions $\vect{G}_\mathrm{S} \left( \vect{r}; \hat{\vect{x}},\hat{\vect{z}} \right)$ and $\vect{G}_\mathrm{S} \left( \vect{r}; \hat{\vect{y}},\hat{\vect{z}} \right)$ are always well defined, also for off-center positioning at $h\neq0$. The cause lies with the symmetrization in Eq.~\eqref{eq:GS}. In combination, as is obvious from  entries 2 and 5 in Tab.~\ref{tab:Gamma_smallK}, the leading order of the symmetrized $\vect{\Gamma}$ shifts to at least order $k^{-1}$, thus ensuring convergence of the integral in Eq.~\eqref{eq:GD_Gamma} around $k=0$. The analogous behavior is observed when considering entries 4 and 6 in Tab.~\ref{tab:Gamma_smallK}. Therefore, the displacements associated with $\vect{G}_\mathrm{S} \left( \vect{r}; \hat{\vect{x}},\hat{\vect{z}} \right)$ and $\vect{G}_\mathrm{S} \left( \vect{r}; \hat{\vect{y}},\hat{\vect{z}} \right)$ remain finite.

The remaining question is, how the diverging displacement fields can arise even in the case of symmetric force dipoles. In fact, {considering elastic films or membranes in three spatial dimensions}, a bending contribution can generally be involved, {as soon as the force dipole is not placed to the center plane at $h=0$}.  
We recall absence of anchoring at infinity. Consequently, we expect bending modes associated with infinite magnitudes of normal displacements at infinite distance from the center of the force dipole.  
Moreover, antisymmetric parts of force dipoles that involve the normal direction lead to overall rotations of the membrane, and therefore again to diverging displacement fields at infinity. Once more, the cause lies with the absence of anchoring of the lateral ends at infinity. {We refer to the remark at the end of Sec.~\ref{sec:convergence-dipoles}. Only for in-plane rotations, we found converging integrals in Eq.~\eqref{eq:GD_Gamma}, implying finite displacements. Thus, again, for the scenario of in-plane rotations, we recover the two-dimensional situation of well-defined displacement fields \cite{richter22}.}

Finally, we remark that in the context of linear elasticity, small strains do not necessarily equate to small displacements. Rather, they signify a small symmetrized gradient in displacement, that is, strain. In other words, we may reside well in the regime of linear elasticity (small strains) but still encounter divergences in the displacement field (for instance, at infinity in an infinitely extended system). Nonlinear elastic response may suppress such divergences.
To our knowledge, there is limited analytical exploration of nonlinear elasticity theory concerning singularities within the framework of free-standing elastic films and membranes. Investigating the challenging nonlinear problem in future research could provide further insight.

\section{Conclusions}
\label{sec:concl}

In summary, we here consider the mechanical displacements and deformations of linearly elastic, initially flat films and membranes exposed to forces and force dipoles. The systems are infinitely extended to the sides, yet of finite, non-vanishing thickness in their normal direction. We address supported films and membranes subject to no-slip boundary conditions on at least one of the two surfaces (top and/or bottom). In these situations, the solutions for resulting displacements in response to both net applied forces and/or stresslets are well defined. For free-standing elastic films and membranes, that is, for stress-free boundary conditions applying to both the top and bottom surface, we observe diverging solutions for most geometries. These divergences can be related to global deformations and large-scale displacements of the whole (infinitely extended) system. When positioning the force dipoles at the center plane of the elastic sheet, our expressions are in line with corresponding considerations for genuinely two-dimensional systems \cite{richter22,lutz22}. 

Our results are significant for any theoretical approach that addresses the deformation of thin, {isotropic} elastic films and membranes of infinite lateral extension {within the linear regime}. Particularly, we find that, unless for very symmetric situations of stresslets included symmetrically at the center plane of the membrane, infinitely extended free-standing membranes cannot serve as immediate theoretical model systems without further precaution. In contrast to genuinely two-dimensional treatments, even off-center stresslets imply divergence of the solutions. Therefore, some stabilization of the system, presumably by introduction of lateral boundaries and/or counterforces, is necessary in this case. 

{
Overall, the analysis in this study relies on linear elasticity, primarily due to the employed formalism of Green's functions and the principle of superposition. Yet, linear elasticity does not always accurately capture the observed response of various materials, particularly in the fields of soft matter and biology. Defining parameters like Young's moduli for biological tissues poses significant challenges, especially when their stiffness is very small in the absence of loading or pre-stretch.
We opt for a linear framework as an initial approach to understand the behavior within the regime of small deformations and to facilitate analytical advancements. Addressing nonlinearities represents an important avenue for future investigations, potentially through perturbative methods. 
Still, we note that the observed divergences in displacement do not necessarily imply that the regime of linear elasticity is violated. Since strains follow as symmetrized gradients in displacements, strains can be small in magnitude even if displacements diverge in infinitely extended systems.
}

{
Finally, it will be interesting to analyze experimental data in view of our results. Particularly, this concerns the analysis and quantification of observed displacement fields. Naturally, real-space experimental systems are always of finite extent, in contrast to our mathematical picture.
It is crucial to acknowledge that the role of thermal fluctuations may become significant with increasing extension. Potentially, they obscure certain aspects of the described effects, especially in large systems. 
Introducing lateral boundaries into our investigations, such as clamping of elastic films and membranes, comparison with experiments is facilitated. 
Mathematically speaking, such additional boundary conditions play a pivotal role. Our study offers valuable insight in this regard, aiming to draw attention to this aspect. We infer that in experiments the role of lateral clamping is dominant for free-standing systems of stress-free top and bottom surfaces, as they are the ones that maintain displacements finite.}
Eventually, if the system is stabilized by a no-slip boundary condition at the bottom surface, while stress-free conditions apply at the top surface, we recover the situation of atomic force microscopy of thin films \cite{kappl2002colloidal,butt2005force}. It is our plan to provide further theoretical tools in this framework.

\begin{acknowledgments}
We acknowledge support of this work by the DFG through the research grant no.\ ME 3571/5-1. A.M.M.\ thanks the Deutsche Forschungsgemeinschaft (DFG, German Research Foundation) for support through the Heisenberg grant no.\ ME 3571/4-1. 
\end{acknowledgments}


\appendix

\setcounter{table}{0}
\renewcommand{\thetable}{A\arabic{table}}

\begin{table*}
	\renewcommand{\arraystretch}{1.4}
	\centering
	\begin{tabular}{|c|c|}
        \hline
		\multicolumn{2}{|c|}{2NOS} \\
		\hline
        \hline
        $D$ & $4\mu k \left( 1+\sigma\right)
		\left( \sigma^2 e^{4k} - 2 \left( \sigma^2+2k^2    \right) e^{2k  } + \sigma^2 \right)$  \\
		\hline
		$a_0^\pm$ & $\pm 2\sigma^2 \left( kh \mp \sigma \right)$ \\
		$a_1^\pm$ & $\sigma \left( k^2  \eta_\pm + 2\sigma k \delta_\pm + 2\sigma^2  \right)$ \\
		~~$a_2^\pm$~~ & ~~~$ -2k^3  \eta_\pm + 4\sigma k^2   \delta_\pm -2\sigma^2 \left( 2 \pm h \right) k + 2\sigma^3 $~~~  \\
		$a_3^\pm$ & $-\sigma \left( k^2    \eta_\mp -2\sigma k \delta_\mp  + 2\sigma^2 \right)$  \\
		\hline
		$b_0^\pm$ & $-2\sigma^2 kh$  \\
		$b_1^\pm$ & $\sigma k \left( 2\sigma h \pm k   \eta_\pm \right)$ \\
		$b_2^\pm$ & $2k \left( \pm k^2 \eta_\pm  - 2\sigma k h    + \sigma^2 h \right)$  \\
		$b_3^\pm$ & $k\sigma \left( \mp k   \eta_\mp - 2\sigma h\right)$  \\
		\hline 
		$g_0^\pm$ & $\mp \sigma^2$  \\
		$g_1^\pm$ & $\pm \sigma \left( k \eta_\pm + \sigma \right)$  \\
		$g_2^\pm$ & $\pm \left( 2k^3 \eta_\pm -\sigma \left( 2k-\sigma \right)   \right)$  \\
		$g_3^\pm$ & $\pm \sigma \left( k \eta_\mp-\sigma  \right)$ \\
		\hline
		$q_0^\pm$ & $\sigma^2 $  \\
		$q_1^\pm$ & $\sigma \left( k\eta_\pm-\sigma  \right) $  \\
		$q_2^\pm$ & $- 2k^2 \eta_\pm -\sigma \left( 2k+\sigma \right)  $  \\
		$q_3^\pm$ & $\sigma \left(  k\eta_\mp + \sigma \right) $  \\
		\hline
        \multicolumn{2}{c}{~} \\
        \hline
        \multicolumn{2}{|c|}{2STF} \\
		\hline
		\hline
		$D$ & $4\mu k \left( 1+\sigma \right)
		\left( e^{4k} - 2\left( 1+2k^2 \right) e^{2k} + 1 \right) $ \\
		\hline
        $a_0^\pm$  & $ 2 \left(  \pm kh -\sigma \right)$ \\
		$a_1^\pm$  & $- \left( k^2 \eta_\pm + 2\sigma k \delta_\pm +1+\sigma^2 \right)$ \\
		~~$a_2^\pm$~~ & ~~$2\left( -k^3\eta_\pm + 2\sigma k^2 \delta_\pm - k \left( \sigma^2 + \delta_\pm \right) + \sigma \right)$~~ \\
		$a_3^\pm$ & $ k^2 \eta_\mp - 2\sigma k \delta_\mp + 1+\sigma^2$ \\
		\hline
		$b_0^\pm$  & $-2kh$ \\
		$b_1^\pm$  & $\mp k^2 \eta_\pm - 2\sigma kh \pm \sigma^2 \mp 1$ \\
  $b_2^\pm$  & $ 2k \left( \pm k^2 \eta_\pm - 2\sigma kh \mp \sigma^2  \pm \delta_\pm \right)$ \\
		$b_3^\pm$ & $\pm k^2 \eta_\mp + 2\sigma kh \mp \sigma^2 \pm 1$ \\
		\hline 
		$g_0^\pm$  & $\mp 1$ \\
		$g_1^\pm$ & $\mp \left(   k\eta_\pm + \sigma \right)$ \\
		$g_2^\pm$ & $\pm \left(   2k^2 \eta_\pm - 2\sigma k + 1\right)$ \\
		$g_3^\pm$ & $\mp \left( k\eta_\mp -\sigma  \right)$ \\
		\hline
		$q_0^\pm$ & $1$ \\
		$q_1^\pm$ & $ \sigma - k \eta_\pm  $ \\
		$q_2^\pm$ & $- \left( 2k^2 \eta_\pm + 2\sigma k + 1\right) $ \\
		$q_3^\pm$ & $ -\left(  k \eta_\mp  + \sigma\right)$ \\
		\hline
	\end{tabular}
 \quad
 \renewcommand{\arraystretch}{1.48}
	\centering
    \begin{tabular}{|c|c|}
		\hline
		\multicolumn{2}{|c|}{NOS-STF} \\
		\hline
        \hline
		$D$ & $4\mu k(1+\sigma) \left( \sigma \left( 1+e^{4k} \right) + \left( 1+\sigma^2+4k^2 \right) e^{2k} \right)$ \\
		\hline
		$a_0^\pm$ & $-2\sigma \left( \sigma \pm kh \right) $ \\
		$a_1^+$ & $ k^2 \eta_+ + 2\sigma k \delta_+ + 2\sigma^2 $ \\
		$a_1^-$ & $-\sigma \left(   k^2 \eta_- + 2k\sigma \delta_- + 1+\sigma^2 \right)$ \\
		$a_2^+$ & $ 2k^3 \eta_+ - 4\sigma k^2 \delta_+ +  \left( 3\sigma^2+\eta_+ \right)k - \sigma^3-\sigma $ \\
		~~$a_2^-$~~ & ~$ 2k^3 \eta_- - 4\sigma k^2 \delta_- + \left( 1+(3-2h)\sigma^2 \right)k - \sigma^3 - \sigma $~ \\
		$a_3^+$ & $\sigma \left( k^2 \eta_- - 2\sigma k \delta_-  + 1+\sigma^2\right)$ \\
		$a_3^-$ & $- k^2 \eta_+ + 2\sigma k \delta_+ -2\sigma^2  $ \\
		\hline
		$b_0^\pm$ & $-2\sigma kh$ \\
		$b_1^+$ & $k \left(  k\eta_+ + 2\sigma h \right) $ \\
		$b_1^-$ & $\sigma \left(k^2 \eta_- - 2\sigma kh + 1 -\sigma^2\right) $ \\
		$b_2^+$ & $-2k^3 \eta_+ + 4\sigma k^2 h + \left( \sigma^2 -\eta_+\right)k+\sigma^3-\sigma $ \\
		$b_2^-$ & $2\eta_-k^3 +4\sigma k^2 h + \left( 1-\eta_+\sigma^2 \right)k + \sigma^3-\sigma $ \\
		$b_3^+$ & $\sigma \left(  k^2 \eta_- +2\sigma kh + 1-\sigma^2 \right) $ \\
		$b_3^-$ & $k \left( k\eta_+ - 2\sigma h \right) $ \\
		\hline
		$g_0^\pm$ & $\mp \sigma $ \\
		$g_1^+$ & $  k\eta_+ + \sigma   $ \\
		$g_1^-$ & $ \sigma \left( k\eta_- + \sigma \right) $ \\
		$g_2^+$ & $ - 2k^2 \eta_+ + \sigma \left( 2k - \sigma  \right)  $ \\
		$g_2^-$ & $  2k^2 \eta_- -2\sigma k + 1   $ \\
		$g_3^+$ & $ \sigma \left( \sigma - k\eta_- \right) $ \\
		$g_3^-$ & $ \sigma - k\eta_+  $ \\
		\hline
		$q_0^\pm$ & $\sigma $ \\
		$q_1^+$ & $  k\eta_+ - \sigma   $ \\
		$q_1^-$ & $ \sigma \left(\sigma -  k\eta_- \right) $ \\
		$q_2^+$ & $  2k^2 \eta_+ + 2\sigma k+\sigma^2   $ \\
		$q_2^-$ & $  2k^2 \eta_- + 2\sigma k+1   $ \\
		$q_3^+$ & $ -\sigma \left( \sigma + k\eta_- \right) $ \\
		$q_3^-$ & $ k\eta_+ + \sigma  $ \\
		\hline
        \multicolumn{2}{c}{~} \\
        \hline
        \multicolumn{2}{|c|}{Definition of coefficients} \\
		\hline
		\hline
            $\sigma$ & $3-4\nu$ \\
            $\delta_\pm$ &  $1 \pm h$ \\
            $\eta_\pm $ & $1 \pm 2h$ \\
            \hline
	\end{tabular}
        \caption{Expressions for $a_n^\pm$, $b_n^\pm$, $g_n^\pm$, and $q_n^\pm$, where $n\in\{0,1,2,3\}$, defining the coefficients $c_2^\pm$ and $c_3^\pm$ {introduced in Eqs.~\eqref{eq:c1c2}}. {These coefficients specify the longitudinal and normal components of the two-dimensionally Fourier-transformed image solutions for the displacement field, $\widetilde{u}_l^*$ and $\widetilde{u}_z^*$, respectively, see Eqs.~\eqref{eq:ul-und-uz_Star} and \eqref{eq:c4c5}. The expressions vary depending on the boundary conditions imposed at the top and bottom surface.} We address geometries of two no-slip boundaries  (2NOS), two  stress-free boundaries (2STF), and the combination of one bottom no-slip and one top stress-free boundary (NOS-STF).
        {For convenience, the definitions of $\sigma$, $\delta_\pm$, and~$\eta_\pm$, see Eq.~(\ref{eq:sigma}) and thereafter, are repeated on the bottom right.}
        }
        \label{tab:coefficients}
\end{table*}

\begin{table*}
    \renewcommand{\arraystretch}{1.9}
    \centering
    \begin{tabular}{|c|c|}
    \hline
    ~~~$\boldsymbol{\Gamma} ( \vect{r}, k ; \hat{\vect{a}}, \hat{\vect{b}} ) $ ~~~ & Exact expression in terms of the monopole solution \\
    \hline
    \hline
    ~~~~$\boldsymbol{\Gamma} ( \vect{r}, k ; \hat{\vect{x}}, \hat{\vect{x}} ) \cdot \hat{\vect{x}}$~~~~ & 
    ~~~~~$k \left( \left( 3\widetilde{\mathcal{G}}_{ll}(k,z) + \widetilde{\mathcal{G}}_{tt}(k,z) \right) J_1(\rho k) \cos\theta + \left( \widetilde{\mathcal{G}}_{tt}(k,z)-\widetilde{\mathcal{G}}_{ll}(k,z) \right) J_3(\rho k) \cos \left( 3\theta\right) \right)$~~~~~ \\
    $\boldsymbol{\Gamma} ( \vect{r}, k ; \hat{\vect{x}}, \hat{\vect{x}} ) \cdot \hat{\vect{y}}$ & $ k\left( \widetilde{\mathcal{G}}_{tt}(k,z) - \widetilde{\mathcal{G}}_{ll}(k,z) \right) \left( J_3(\rho k) \sin(3\theta) - J_1(\rho k) \sin\theta \right) $ \\
    $\boldsymbol{\Gamma} ( \vect{r}, k ; \hat{\vect{x}}, \hat{\vect{x}} ) \cdot \hat{\vect{z}}$ & $ 2i k \, \widetilde{\mathcal{G}}_{zl}(k,z) \left(  J_2(\rho k) \cos(2\theta) - J_0(\rho k) \right) $ \\
    \hline
    $\boldsymbol{\Gamma} ( \vect{r}, k ; \hat{\vect{y}}, \hat{\vect{x}} ) \cdot \hat{\vect{x}}$ & $k\left( \widetilde{\mathcal{G}}_{tt}(k,z)-\widetilde{\mathcal{G}}_{ll}(k,z) \right) \left( J_3(\rho k) \sin(3\theta)-J_1(\rho k) \sin\theta \right)$ \\
    $\boldsymbol{\Gamma} ( \vect{r}, k ; \hat{\vect{y}}, \hat{\vect{x}} ) \cdot \hat{\vect{y}}$ & $k \left(\left( 3\widetilde{\mathcal{G}}_{tt}(k,z) + \widetilde{\mathcal{G}}_{ll}(k,z) \right) J_1(\rho k) \cos\theta - \left( \widetilde{\mathcal{G}}_{tt}(k,z)-\widetilde{\mathcal{G}}_{ll}(k,z) \right) J_3(\rho k) \cos \left( 3\theta\right) \right)$ \\
    $\boldsymbol{\Gamma} ( \vect{r}, k ; \hat{\vect{y}}, \hat{\vect{x}} ) \cdot \hat{\vect{z}}$ & $2ik \, \widetilde{\mathcal{G}}_{zl}(k,z) J_2(\rho k) \sin(2\theta)$ \\
    \hline
    $\boldsymbol{\Gamma} ( \vect{r}, k ; \hat{\vect{z}}, \hat{\vect{x}} ) \cdot \hat{\vect{x}}$ & $2ik \, \widetilde{\mathcal{G}}_{lz}(k,z) \left( J_2(\rho k) \cos(2\theta)-J_0(\rho k) \right) $  \\
    $\boldsymbol{\Gamma} ( \vect{r}, k ; \hat{\vect{z}}, \hat{\vect{x}} ) \cdot \hat{\vect{y}}$ & $2ik \, \widetilde{\mathcal{G}}_{lz}(k,z) J_2(\rho k) \sin(2\theta)$ \\
    $\boldsymbol{\Gamma} ( \vect{r}, k ; \hat{\vect{z}}, \hat{\vect{x}} ) \cdot \hat{\vect{z}}$ & $4 k\, \widetilde{\mathcal{G}}_{zz}(k,z) J_1(\rho k) \cos\theta $ \\
    \hline
    \hline
    $\boldsymbol{\Gamma} ( \vect{r}, k ; \hat{\vect{x}}, \hat{\vect{y}} ) \cdot \hat{\vect{x}}$ & $k \left( \left( 3\widetilde{\mathcal{G}}_{tt}(k,z) + \widetilde{\mathcal{G}}_{ll}(k,z) \right) J_1(\rho k) \sin\theta + \left( \widetilde{\mathcal{G}}_{tt}(k,z)-\widetilde{\mathcal{G}}_{ll}(k,z) \right) J_3(\rho k) \sin \left( 3\theta\right) \right)$ \\
    $\boldsymbol{\Gamma} ( \vect{r}, k ; \hat{\vect{x}}, \hat{\vect{y}} ) \cdot \hat{\vect{y}}$ & $-k\left( \widetilde{\mathcal{G}}_{tt}(k,z)-\widetilde{\mathcal{G}}_{ll}(k,z) \right) \left( J_1(\rho k) \cos\theta + J_3(\rho k) \cos(3\theta) \right)$ \\
    $\boldsymbol{\Gamma} ( \vect{r}, k ; \hat{\vect{x}}, \hat{\vect{y}} ) \cdot \hat{\vect{z}}$ & $2ik \, \widetilde{\mathcal{G}}_{zl}(k,z) J_2(\rho k) \sin(2\theta)$ \\
    \hline
    $\boldsymbol{\Gamma} ( \vect{r}, k ; \hat{\vect{y}}, \hat{\vect{y}} ) \cdot \hat{\vect{x}}$ & $-k\left( \widetilde{\mathcal{G}}_{tt}(k,z)-\widetilde{\mathcal{G}}_{ll}(k,z) \right) \left( J_1(\rho k) \cos\theta + J_3(\rho k) \cos(3\theta) \right)$ \\
    $\boldsymbol{\Gamma} ( \vect{r}, k ; \hat{\vect{y}}, \hat{\vect{y}} ) \cdot \hat{\vect{y}}$ & $k \left( \left( 3\widetilde{\mathcal{G}}_{ll}(k,z) + \widetilde{\mathcal{G}}_{tt}(k,z)\right) J_1(\rho k) \sin\theta - \left( \widetilde{\mathcal{G}}_{tt}(k,z)-\widetilde{\mathcal{G}}_{ll}(k,z) \right) J_3(\rho k) \sin(3\theta)  \right) $ \\
    $\boldsymbol{\Gamma} ( \vect{r}, k ; \hat{\vect{y}}, \hat{\vect{y}} ) \cdot \hat{\vect{z}}$ & $2ik \, \widetilde{\mathcal{G}}_{zl}(k,z) \left( J_0(\rho k) + J_2(\rho k) \cos(2\theta)\right)$ \\
    \hline
    $\boldsymbol{\Gamma} ( \vect{r}, k ; \hat{\vect{z}}, \hat{\vect{y}} ) \cdot \hat{\vect{x}}$ & $2ik \, \widetilde{\mathcal{G}}_{lz}(k,z) J_2(\rho k) \sin(2\theta)$ \\
    $\boldsymbol{\Gamma} ( \vect{r}, k ; \hat{\vect{z}}, \hat{\vect{y}} ) \cdot \hat{\vect{y}}$ & $-2ik \, \widetilde{\mathcal{G}}_{lz}(k,z) \left( J_0(\rho k) + J_2(\rho k) \cos(2\theta) \right)$ \\
    $\boldsymbol{\Gamma} ( \vect{r}, k ; \hat{\vect{z}}, \hat{\vect{y}} ) \cdot \hat{\vect{z}}$ & $4k\, \widetilde{\mathcal{G}}_{zz}(k,z) J_1(\rho k) \sin\theta$ \\
    \hline
    \hline
    $\boldsymbol{\Gamma} ( \vect{r}, k ; \hat{\vect{x}}, \hat{\vect{z}} ) \cdot \hat{\vect{x}}$ & $2 \left( \partial_h \left( \widetilde{\mathcal{G}}_{tt}(k,z) + \widetilde{\mathcal{G}}_{ll}(k,z) \right) J_0(\rho k) + 
    \partial_h \left( \widetilde{\mathcal{G}}_{tt}(k,z) - \widetilde{\mathcal{G}}_{ll}(k,z) \right) J_2(\rho k) \cos(2\theta)
    \right) $ \\
    $\boldsymbol{\Gamma} ( \vect{r}, k ; \hat{\vect{x}}, \hat{\vect{z}} ) \cdot \hat{\vect{y}}$ & $2\partial_h \left( \widetilde{\mathcal{G}}_{tt}(k,z) - \widetilde{\mathcal{G}}_{ll}(k,z) \right) J_2(\rho k) \sin(2\theta)$ \\
    $\boldsymbol{\Gamma} ( \vect{r}, k ; \hat{\vect{x}}, \hat{\vect{z}} ) \cdot \hat{\vect{z}}$ & $4i \, \partial_h \widetilde{\mathcal{G}}_{zl}(k,z) J_1(\rho k) \cos\theta $ \\
    \hline
    $\boldsymbol{\Gamma} ( \vect{r}, k ; \hat{\vect{y}}, \hat{\vect{z}} ) \cdot \hat{\vect{x}}$ & $2\partial_h \left( \widetilde{\mathcal{G}}_{tt}(k,z) - \widetilde{\mathcal{G}}_{ll}(k,z) \right) J_2(\rho k) \sin(2\theta)$ \\
    $\boldsymbol{\Gamma} ( \vect{r}, k ; \hat{\vect{y}}, \hat{\vect{z}} ) \cdot \hat{\vect{y}}$ & $2 \left( \partial_h \left( \widetilde{\mathcal{G}}_{tt}(k,z) + \widetilde{\mathcal{G}}_{ll}(k,z) \right) J_0(\rho k) - 
    \partial_h \left( \widetilde{\mathcal{G}}_{tt}(k,z) - \widetilde{\mathcal{G}}_{ll}(k,z) \right) J_2(\rho k) \cos(2\theta)
    \right) $  \\
    $\boldsymbol{\Gamma} ( \vect{r}, k ; \hat{\vect{y}}, \hat{\vect{z}} ) \cdot \hat{\vect{z}}$ & $4i \, \partial_h \widetilde{\mathcal{G}}_{zl}(k,z) J_1(\rho k) \sin\theta $ \\
    \hline
    $\boldsymbol{\Gamma} ( \vect{r}, k ; \hat{\vect{z}}, \hat{\vect{z}} ) \cdot \hat{\vect{x}}$ & $4i \, \partial_h \widetilde{\mathcal{G}}_{lz}(k,z) J_1(\rho k) \cos\theta$ \\
    $\boldsymbol{\Gamma} ( \vect{r}, k ; \hat{\vect{z}}, \hat{\vect{z}} ) \cdot \hat{\vect{y}}$ & $4i \, \partial_h \widetilde{\mathcal{G}}_{lz}(k,z) J_1(\rho k) \sin\theta$ \\
    $\boldsymbol{\Gamma} ( \vect{r}, k ; \hat{\vect{z}}, \hat{\vect{z}} ) \cdot \hat{\vect{z}}$ & $4 \, \partial_h \widetilde{\mathcal{G}}_{zz}(k,z) J_0(\rho k) $ \\
    \hline
\end{tabular}
    \caption{Expressions for the various combinations of the function { $\boldsymbol{\Gamma}(\vect{r},k;\hat{\vect{a}},\hat{\vect{b}})$, where $\hat{\vect{a}},\hat{\vect{b}}  \in\{\hat{\vect{x}},\hat{\vect{y}},\hat{\vect{z}}\} $, 
    are listed componentwise by projecting them for each combination on the Cartesian unit vectors. }
    The function $\vect{\Gamma}$ upon integration defines the Green's function of the force dipole in real space, see Eq.~\eqref{eq:GD_Gamma}. We here express $\vect{\Gamma}$ in the system of polar coordinates $(\rho, \theta)$.
    In these expressions, $ \widetilde{ \mathcal{G} }_{ij} $ represents the two-dimensionally Fourier-transformed Green's function quantifying the displacements induced in the elastic system in response to a force acting at position $(0,0,h)$ on the elastic film or membrane of finite thickness, {see Eqs.~\eqref{eq:Gsum}--\eqref{eq:A}.}
    }
    \label{tab:Gamma}
\end{table*}

\section{Appendix: Tables of expressions}
\label{appendix:tables}

In this Appendix, we present tables containing the remaining expressions that specify the Green's functions introduced in the main text. While not essential for comprehending the contents, they detail the Green's functions for reference.
In Tab.~\ref{tab:coefficients}, we include the expressions for $D$, $a_n^\pm$, $b_n^\pm$, $g_n^\pm$, and $q_n^\pm$, $n\in\{0,1,2,3\}$, as defined in Eqs.~\eqref{eq:c1c2}. They determine the coefficients $c_2^\pm$ and $c_3^\pm$ {introduced in Eqs.~\eqref{eq:ul-und-uz_Star} and \eqref{eq:c4c5}, thus governing the image solution for the longitudinal and normal components of the displacement field. Consequently, they also enter the corresponding part of the Green's function in Eqs.~\eqref{eq:Gij_Star} and \eqref{eq:A}.}
The expressions depend on the specific set of boundary conditions and are tabulated accordingly. {Moreover, Tab.~\ref{tab:Gamma} includes expressions of the Cartesian components of the function $\vect{\Gamma}(\vect{r},k;\hat{\vect{a}},\hat{\vect{b}})$, $\hat{\vect{a}},\hat{\vect{b}}\in\{\hat{\vect{x}},\hat{\vect{y}},\hat{\vect{z}}\}$. Via Eq.~\eqref{eq:GD_Gamma}, this function determines the} Green's function for the force dipole $\vect{G}_\mathrm{D} (\vect{r}; \hat{\vect{a}}, \hat{\vect{b}})$. 

\section{{Two-dimensional inverse Fourier transformation in polar coordinates}}
\label{appendix:fourier}

{
In this Appendix, we summarize expressions for calculating two-dimensional inverse Fourier transformations in polar coordinates. For a comprehensive cover and associated proofs, classical textbooks on Fourier analysis~\cite{bracewell99, folland2009fourier, stein2011fourier} offer detailed insight. Additionally, a concise summary of the methods is found in Ref.~\onlinecite{baddour2011two}.}

{
The two-dimensional inverse Fourier transformation of a given function expressed in polar coordinates as $\widetilde{f}(k,\phi)$ can be represented through Fourier series as
\begin{equation}
    f(\rho,\theta) = 
    \mathscr{F}^{-1} \left\{ \widetilde{f} (k, \phi) \right\} 
    = \sum_{n=-\infty}^\infty f_n(\rho) \, e^{in\theta} \, , 
    \label{eq:infFourier_appendix}
\end{equation}
where
\begin{equation}
    f_n(\rho) = \frac{i^n}{2\pi} \int_0^\infty 
    \widetilde{f}_n(k) J_n(\rho k) \, k \, \Intd k \, , 
\end{equation}
with
\begin{equation}
    \widetilde{f}_n(k) = \frac{1}{2\pi}
    \int_0^{2\pi} \widetilde{f}(k,\phi) \, e^{-in\phi} \, 
    \Intd \phi \, . \label{eq:fn_appendix}
\end{equation}
Here, again,  $J_n$ refers to the Bessel function of the first kind of order $n$~\cite{abramowitz72}.
}

{
When dealing with a radially symmetric function $\widetilde{f} (k)$ independent of the azimuthal angle $\phi$, only the term $n=0$ persists in the series described by Eq.~\eqref{eq:infFourier_appendix}.
As a result, the inverse Fourier transformation simplifies to
\begin{equation}
    f(\rho) = 
    \mathscr{F}^{-1} \left\{ \widetilde{f} (k) \right\}
    = \frac{1}{2\pi} \int_0^\infty 
    \widetilde{f}(k) J_0(\rho k) \, k \, \Intd k \, .
\end{equation}
This specific formulation applies to the component $\mathcal{G}_{zz}$ of the Green's function, as indicated by Eq.~\eqref{eq:Gzz}.
For the other components, we observe that only terms of $n \in \{0,1,2\}$ remain for the force monopole and terms of $n \in \{0,1,2,3\}$ for the force dipole.
The entries of Tab.~\ref{tab:Gamma} were derived using Eqs.~\eqref{eq:infFourier_appendix}-- \eqref{eq:fn_appendix}.
}

%

\end{document}